\documentclass[12pt,nofootinbib]{revtex4}
\usepackage{amssymb}
\usepackage{amsmath, graphicx}
\usepackage{slashed}
\usepackage[unicode]{hyperref}
\usepackage{csquotes}
\usepackage{braket}
\usepackage{natbib}
\usepackage{array}
\usepackage{xcolor}

\begin{document}
\title{Constraining regular primordial black holes with isocurvature gravitational waves}
\author{Ngo Phuc Duc Loc}
\thanks{Email: locngo148@gmail.com}

\affiliation{Physics Division, National Center for Theoretical Sciences,
National Taiwan University, Taipei 106319, Taiwan}

\begin{abstract}
We find the constraint on the population of ultra-light regular primordial black holes (RPBHs) by using isocurvature gravitational waves (GW). If ultra-light RPBHs dominated the early Universe, the initial isocurvature perturbation is converted into curvature perturbation that induce second-order GW background upon evaporation of RPBHs. The upper limit of extra relativistic degrees of freedom $\Delta N_{\rm eff}$, which could be inferred from Big Bang Nucleosynthesis or Cosmic Microwave Background observations, places a constraint on the maximum energy density of GW, which in turn can be used to constrain the RPBH population. As RPBHs have different lifetime from their singular counterparts, the constraint must be modified accordingly. While the formalism that we provide is generic, we work out explicitly the case of Simpson-Visser metric for demonstration. The RPBHs associated with this metric have lower Hawking temperature and smaller horizon size, leading to a longer lifetime than the singular Schwarzschild black holes. This implies a stronger constraint on RPBH population as they dominate the Universe for a longer period of time and generate stronger GW.
\end{abstract}

\maketitle

\section{Introduction}
The existence of singularity at the center of the usual black hole (BH) solutions, where energy density and curvature of spacetime are infinite, is a problematic issue. It indicates that our understanding of BH's interior is incomplete and perhaps new physics is needed in order to avoid such a singularity. A variety of regular black hole (RBH) models have been proposed, ranging from phenomenological metrics \cite{Bardeen,Hayward:2005gi,Dymnikova:1992ux,Culetu:2013fsa,Culetu:2014lca,Ghosh:2014pba,Simpson:2019mud,Simpson:2018tsi} to quantum gravity-inspired metrics \cite{Peltola:2008pa,Peltola:2009jm,Bianchi:2018mml,DAmbrosio:2018wgv,Zhang:2024khj}. Recent advancements in the study of dynamical formation of RBH \cite{Bueno:2024dgm,Bueno:2024eig,Bueno:2024zsx,Bueno:2025zaj} have also received great interests.

From the perspective of cosmology, it would be interesting to explore how RBHs could have observational impacts. A prominent case stands out if RBHs have  primordial origin, dubbed regular primordial black holes (RPBHs). RPBHs have the same motivations as the singular primordial black holes (PBHs) and, in addition, are singularity-free. RPBHs as a dark matter (DM) candidate has been explored in recent works \cite{Calza:2024fzo,Calza:2024xdh,Calza:2025mwn}. For non-self-similar RPBHs, their remnant states could also be a DM candidate \cite{Davies:2024ysj,Trivedi:2025vry,Ong:2024dnr,Dymnikova:2015yma} though this possibility has also been questioned \cite{Khodadi:2025icd}. If RPBHs are instead self-similar, they evaporate completely and DM can be produced from their evaporation \cite{Loc:2025mzc}. In these studies, it has been shown that the parameter space could be shifted by orders of magnitude, opening up a compelling potential to ``probe" BH's interior by means of its modified evaporation dynamics.

For self-similar ultra-light RPBHs, their evaporation products thermalize with the background plasma, so the only viable way to probe this kind of BHs is to study their gravitational wave (GW) signatures. There are many potential sources of GW from RPBHs' formation, merger, evaporation. In this paper, we want to study specifically the isocurvature GW signal of ultra-light RPBHs. It has been known that if ultra-light PBHs temporarily dominated the early Universe, the initial isocurvature perturbation, which is basically the relative fluctuations between PBH and radiation background, is converted into curvature perturbation that induces second-order GW background after evaporation of PBH \cite{Papanikolaou:2020qtd,Domenech:2020ssp,Domenech:2021wkk,Inomata:2020lmk,Papanikolaou:2022chm}. This kind of GW background could be very strong and may potentially violate the $\Delta N_\text{eff}$ bound on extra relativistic degrees of freedom. This implies that we can place an upper bound on the PBH population such that the total GW signal they produce can still be consistent with observation. Here, we generalize this idea to the case of RPBHs. As RPBHs have distinct lifetime from the singular Schwarzschild PBH, the GW signal and hence the constraint could be shifted by orders of magnitude. While the machinery that we provide is generic, we demonstrate the idea with the Simpson-Visser metric which is usually regarded as a minimal extension of the singular Schwarzschild metric. This metric leads to an extended lifetime of RPBH, resulting in stronger GW signal and thus stricter bound on RPBH population.

This paper is organized as follows. After reviewing the evaporation of RPBHs in Sec. \ref{sec:Evaporation of RPBH}, we obtain formulae for isocurvature GW signal and the constraint on RPBH population in Sec. \ref{seC:Isocurvature GW}. We present our results in Sec. \ref{sec:Results} and then conclude in Sec. \ref{sec:Conclusion}. We use the natural units in which $\hbar=c=k_B=1$.

\section{Evaporation of regular primordial black holes}\label{sec:Evaporation of RPBH}

We first review the evaporation dynamics of RPBHs \cite{Loc:2025mzc}. We consider a class of static, spherically symmetric metrics of the form:
\begin{equation}
    ds^2=-f(r)dt^2+\frac{dr^2}{g(r)}+h(r)d\Omega^2,
\end{equation}
where $d\Omega^2$ is the usual 2-sphere metric. The Hawking radiation is defined as \cite{Hawking:1974rv,Hawking:1975vcx,Gibbons:1977mu}:
\begin{equation}
    T_\text{H}=\frac{\kappa}{2\pi}=\frac{f'(r)}{4\pi}\sqrt{\frac{g(r)}{f(r)}}\Bigg|_{r=r_\text{H}},
\end{equation}
where $\kappa$ is the surface gravity, prime denotes derivative with respect to $r$, and $r_\text{H}$ is the outer horizon radius and is identified as the largest root of the equation $f(r)=0$. We define the ratios:
\begin{equation}
    A(l)\equiv\frac{T_\text{H}}{T_\text{Sch}}\hspace{1cm},\hspace{1cm}
    B(l)\equiv\frac{r_\text{H}}{r_\text{Sch}}.
\end{equation}
Here, $l$ is the dimensionless regularizing parameter introduced to avoid the appearance of singularity. The explicit functional forms of $A(l)$ and $B(l)$ depend on the chosen metric (see Fig. 2 of \cite{Loc:2025mzc} for more details). The subscript ``Sch" denotes quantities of singular Schwarzschild BH and are given by
\begin{equation}
    T_\text{Sch}=\frac{1}{8\pi G M}\hspace{1cm},\hspace{1cm}r_\text{Sch}=2GM.
\end{equation}
Here, $M$ is the BH's mass and $G$ is the Newtonian gravitational constant (in natural units, we have $G=1/m_\text{pl}^2$ with $m_\text{pl}$ is the Planck mass). The lifetime of RPBH is given by:
\begin{equation}\label{eq:lifetime}
    \tau_\text{RPBH}=\frac{1}{B(l)^2A(l)^4}\frac{10240\pi M_i^3}{g_*(T_\text{H})m_\text{pl}^4},
\end{equation}
where $g_*(T_\text{H})$ is the number of relativistic degrees of freedom (d.o.f.) at temperature $T_\text{H}$ and $M_i$ is the initial mass of RPBH. The cosmic temperature at which the RPBHs evaporate is
\begin{equation}\label{eq:T_eva}
    T_\text{eva}\simeq A(l)^2B(l)\frac{\sqrt{3}}{64\times 5^{1/4}}\frac{g_*(T_\text{H})^{1/2}m_\text{pl}^{5/2}}{\pi^{5/4}g_{*,\text{eva}}^{1/4}M_i^{3/2}},
\end{equation}
where $g_{*,\text{eva}}$ is the number of relativistic d.o.f. at the evaporation time.

The initial mass of RPBH at the formation time is given by:
\begin{equation}\label{eq:Mi}
    M_i=\gamma\frac{m_\text{pl}^2}{2H_i},
\end{equation}
where $H_i$ is the Hubble rate at the formation time and we choose $\gamma\sim 0.1$ in the following. In a standard radiation-dominated (RD) Universe before Big Bang Nucleosynthesis (BBN), the initial fraction of total energy density in RPBHs is defined as $\beta=\rho_\text{RPBH,i}/\rho_\text{rad,i}$, where $\rho_\text{RPBH,i}$ is the initial energy density of RPBH and $\rho_\text{rad,i}$ is the initial energy density of radiation. The term ``population" is also commonly used to refer to $\beta$. As RPBHs behave like matter, their energy density is diluted slower than radiation. Therefore, there exists a critical value of the population above which there is RPBH domination:
\begin{equation}\label{eq:beta_c}
    \beta_c=1.78\times 10^{-6}\gamma^{-1/2}A(l)^2B(l)\left(\frac{g_*(T_\text{H})}{106.75}\right)^{1/2}\left(\frac{106.75}{g_{*\text{,eva}}}\right)^{1/4}\left(\frac{g_{*,i}}{106.75}\right)^{1/4}\left(\frac{\text{g}}{M_i}\right),
\end{equation}
where $g_{*,i}$ is the number of relativistic d.o.f. at the formation time.

\section{Isocurvature gravitational waves and $\Delta N_\text{eff}$ constraint}\label{seC:Isocurvature GW}

We consider a gas of RPBHs following Poisson statistics, meaning that they are evenly distributed in space. The mean  physical distance between nearby RPBHs at the formation time is then \cite{Papanikolaou:2020qtd,Domenech:2020ssp}:
\begin{equation}
    d_i=\left(\frac{3M_i}{4\pi\rho_\text{RPBH,i}}\right)^{1/3}=\gamma^{1/3}\beta^{-1/3}H_i^{-1},
\end{equation}
where we used Eq. \ref{eq:Mi} and the standard Friedmann equation for a RD Universe $\rho_\text{rad,i}=3H_i^2m_\text{pl}^2/8\pi$. Thus, there exists a smallest scale below which the continuous fluid description of RPBHs is no longer valid:
\begin{equation}\label{eq:kUV}
    k_\text{UV}=\frac{a_i}{d_i}=a_i\gamma^{-1/3}\beta^{1/3}H_i.
\end{equation}
Here, $a_i$ is the cosmological scale factor at the RPBH formation time. We also define $k_i,k_\text{eq},k_\text{eva}$ to be the comoving wavenumber at RPBH formation, RPBH-radiation equality, and RPBH's evaporation respectively. We will find it useful to obtain the ratios of these scales as follows:
\begin{equation}\label{eq:keq/ki}
    \frac{k_\text{eq}}{k_i}=\frac{a_\text{eq}H_\text{eq}}{a_iH_i}=\frac{a_\text{eq}}{a_i}\sqrt{2}\left(\frac{T_\text{eq}}{T_i}\right)^2=\sqrt{2}\beta,
\end{equation}
where we used the fact that the Hubble rate at the equality time is  $H_\text{eq}\propto\sqrt{2}T_\text{eq}^2$, the Hubble rate at the formation time is $H_i\propto T_i^2$, and $\rho_\text{RPBH}/\rho_\text{rad}\propto a\Rightarrow \beta=a_i/a_\text{eq}=T_\text{eq}/T_i$. Here, $a_i,a_\text{eq}$ are scale factors and $T_i,T_\text{eq}$ are cosmic temperatures at the RPBH formation time and at the RPBH-radiation equality time respectively.  From Eq. \ref{eq:kUV}, we have
\begin{equation}\label{eq:kUV/ki}
    \frac{k_\text{UV}}{k_i}=\gamma^{-1/3}\beta^{1/3}.
\end{equation}
From Eqs. \ref{eq:keq/ki} and \ref{eq:kUV/ki}, we obtain
\begin{equation}\label{eq:keq/kUV}
    \frac{k_\text{eq}}{k_\text{UV}}=\sqrt{2}\gamma^{1/3}\beta^{2/3}.
\end{equation}
Next, we have 
\begin{align}
    \frac{k_\text{eva}}{k_i}&=\frac{a_\text{eva}H_\text{eva}}{a_iH_i}\\
    &=\left(\frac{n_\text{RPBH,i}}{n_\text{RPBH,eva}}\right)^{1/3}\frac{H_\text{eva}}{H_i}\\
    &\simeq \beta^{1/3}\left(\frac{\rho_\text{rad,i}}{\rho_\text{RPBH,eva}}\right)^{1/3}\frac{H_\text{eva}}{H_i}\\
    &=\beta^{1/3}\left(\frac{H_\text{eva}}{H_i}\right)^{1/3}\\
    &=\beta^{1/3}\left(\frac{T_\text{eva}}{T_i}\right)^{2/3}\\
    &=\beta^{1/3}\beta_c^{2/3},\label{eq:keva/ki}
    \end{align}
where the subscript ``eva" denotes quantities at the RPBH's evaporation time. We used the facts that RPBH number density goes like $n_\text{RPBH}\propto a^{-3}$, the mass of RPBH is approximately constant until evaporation, and  $\beta_c=T_\text{eva}/T_i$ \cite{Loc:2025mzc}. From Eqs. \ref{eq:kUV/ki} and \ref{eq:keva/ki}, we obtain
\begin{equation}\label{eq:kUV/keva}
    \frac{k_\text{UV}}{k_\text{eva}}=\gamma^{-1/3}\beta_c^{-2/3}.
\end{equation}
Since entropy is conserved from the RPBH's evaporation time to today, we have
\begin{align}
        k_\text{eva}&=a_\text{eva}H_\text{eva}\\
        &=\frac{a_\text{eva}}{a_0}H_\text{eva}\\
        &\simeq \frac{g_{s,0}^{1/3}T_0}{g_{\text{s,eva}}^{1/3}T_\text{eva}}\Gamma_\text{RPBH}\\
        &=\frac{5^{1/4}\pi^{1/4}}{160\sqrt{3}}\frac{g_{s,0}^{1/3}g_*(T_\text{H})^{1/2}}{g_{*,\text{eva}}^{1/12}}A(l)^2B(l)\left(\frac{m_\text{pl}}{M_i}\right)^{3/2}T_0,
\end{align}
where $g_{s,0}=3.94$ is the number of entropy d.o.f. today, $T_0\approx 2.3\times 10^{-13}\ \rm GeV$ is the photon temperature today, $a_0=1$ is the scale factor today, at high temperature we have $g_{*,\text{eva}}\sim g_\text{s,eva}$, $\Gamma_\text{RPBH}=\tau_\text{RPBH}^{-1}\sim H_\text{eva}$ is the decay rate of RPBH given in Eq. \ref{eq:lifetime}, and $T_\text{eva}$ is given in Eq. \ref{eq:T_eva}. Using $m_\text{pl}\approx 1.22\times 10^{19} \ \rm GeV\approx 2.18\times 10^{-5}\ \rm g$, this can be written in the normalized form as:
\begin{equation}\label{eq:keva}
    \frac{k_\text{eva}}{\text{Mpc}^{-1}}\simeq 2.9\times 10^{17}A(l)^2B(l)\left(\frac{106.75}{g_{*\text{,eva}}}\right)^{1/12}\left(\frac{g_*(T_\textbf{H})}{106.75}\right)^{1/2}\left(\frac{\rm g}{M_i}\right)^{3/2}.
\end{equation}
By using Eqs. \ref{eq:kUV/keva} and \ref{eq:keva}, we get
\begin{equation}\label{eq:kUV vs Mi}
    \frac{k_\text{UV}}{\text{Mpc}^{-1}}\simeq 1.97\times 10^{21}A(l)^{2/3}B(l)^{1/3}\left(\frac{g_*(T_\text{H})}{106.75}\right)^{1/6}\left(\frac{106.75}{g_{*\text{,eva}}}\right)^{-1/12}\left(\frac{g_{*\text{,i}}}{106.75}\right)^{-1/6}\left(\frac{\rm g}{M_i}\right)^{5/6},
\end{equation}
where $\beta_c$ in Eq. \ref{eq:beta_c} has been used.

The isocurvature GW spectrum at the production time is \cite{Papanikolaou:2020qtd,Domenech:2020ssp,Domenech:2021wkk,Domenech:2023jve}
\begin{equation}
    \Omega_\text{GW,c}(k)\simeq 5.94\times 10^{-6}\left(\frac{k_\text{eq}}{k_\text{UV}}\right)^8\left(\frac{k_\text{UV}}{k_\text{eva}}\right)^2\left(\frac{k}{k_\text{eva}}\right)^{11/3}\Theta(k_\text{UV}-k),
\end{equation}
where $\Theta$ is the usual Heaviside theta function \footnote{Instead of a step function, one can also use the following integral prescription for a smoother spectrum around the peak \cite{Domenech:2020ssp}: $\Theta\rightarrow2\int_{-s_0(k)}^{s_0(k)}(1-s^2)^2ds/(1-c_s^2s^2)^{5/3}$, where
$$
s_0(k)=
\begin{cases}
    1\hspace{1cm}\frac{k}{k_\text{UV}}\leq \frac{2}{1+c_s^{-1}}\\
    2\frac{k_\text{UV}}{k}-c_s^{-1}\hspace{1cm}\frac{2}{1+c_s^{-1}}\leq\frac{k}{k_\text{UV}}\leq \frac{2}{c_s^{-1}}\\
    0\hspace{1cm}\frac{k}{k_\text{UV}}\geq \frac{2}{c_s^{-1}}
\end{cases}
$$
with $c_s=1/\sqrt{3}$ is the sound speed of radiation.
}. As mentioned earlier, this cutoff was introduced as RPBH can no longer be treated as a point-mass at very small scales. The subscript ``c" indicates the time where the tensor modes propagate as waves and hence GW spectral density remains constant, which can be identified as the evaporation time of RPBH \cite{Domenech:2020ssp}. We note that this is a general formula applicable for any kind of PBH following Poisson distribution. The difference in the result between singular and regular BH lies at the modifications of the wavenumbers originated from modified evaporation time scale. The GW spectrum redshifted to today is (e.g., see Appendix A of \cite{Loc:2024qbz})
\begin{align}
        \Omega_\text{GW,0}(k)&\approx 3.33\times 10^{-5}\left(\frac{106.75}{g_{*,\text{eva}}}\right)^{1/3}\Omega_\text{GW,c}(k)\\
        &=1.98\times 10^{-10}\left(\frac{106.75}{g_{*,\text{eva}}}\right)^{1/3}\left(\frac{k_\text{eq}}{k_\text{UV}}\right)^8\left(\frac{k_\text{UV}}{k_\text{eva}}\right)^2\left(\frac{k}{k_\text{eva}}\right)^{11/3}\Theta(k_\text{UV}-k).
\end{align}
By using Eqs. \ref{eq:keq/kUV}, \ref{eq:kUV/keva}, \ref{eq:keva}, we can conveniently express this in terms of RPBH mass and population as
\begin{equation}\label{eq:OmegaGW0}
\begin{aligned}
    \Omega_\text{GW,0}(k)&\approx 1.37\times 10^{-65}\gamma^{8/3}A(l)^{-10}B(l)^{-5}\left(\frac{106.75}{g_{*,\text{eva}}}\right)^{-11/36}\left(\frac{g_*(T_\text{H})}{106.75}\right)^{-5/2}\left(\frac{g_{*,i}}{106.75}\right)^{-1/3}\times\\
    &\times\left(\frac{\rm g}{M_i}\right)^{-41/6}\beta^{16/3}\left(\frac{k}{\text{Mpc}^{-1}}\right)^{11/3}\Theta(k_\text{UV}-k).
    \end{aligned}
\end{equation}
Because the scale factor today is set to unity, $a_0=1$, the comoving wavenumber $k$ is also the physical wavenumber today. It is also useful to use the relation
\begin{equation}
    \left(\frac{f}{\rm Hz}\right)=1.55\times 10^{-15}\left(\frac{k}{\rm Mpc^{-1}}\right)
\end{equation}
to convert wavenumber into frequency and vice versa.

GWs behave like radiation (commonly known as ``dark radiation" since they only interact gravitationally with visible matter), so their energy density is constrained by the nondetection of extra relativistic d.o.f. . From the BBN time to the recombination time when the Cosmic Microwave Background (CMB) is produced, the Standard Model predicts only two relativistic species which are photons and neutrinos \footnote{At present, only photons are still relativistic.}:
\begin{align}
        \rho_\text{rad}&=\rho_\gamma+\rho_\nu\\
        &=\rho_\gamma\left(1+\frac{7}{8}\left(\frac{4}{11}\right)^{4/3}N_\text{eff}\right)\\
        &=\rho_\gamma\left(1+\frac{7}{8}\left(\frac{4}{11}\right)^{4/3}(N_\text{eff}^\text{SM}+\Delta N_\text{eff})\right)\\
        &=\rho_\text{rad}^\text{SM}+\frac{7}{8}\left(\frac{4}{11}\right)^{4/3}\Delta N_\text{eff}\ \rho_\gamma,
\end{align}
where $\rho_\gamma$ and $\rho_\nu$ are energy densities of photons and neutrinos respectively, $N_\text{eff}^\text{SM}=3.046$ is the effective number of neutrino species\footnote{It is not exactly 3 due to the correction of $e^+e^-$ annihilation.} and $\Delta N_\text{eff}$ encodes any ``extra" relativistic species. $\rho_\text{rad}^\text{SM}$ is all known visible radiation in the Standard Model, which includes photons and neutrinos. The factor $7/8$ comes from the fermionic statistic of neutrino and the factor $(4/11)^{4/3}$ comes from the fact that photon is slightly hotter than neutrino due to the heating of electron-positron annihilation occurring around the BBN time. If GWs comprise \textit{all} dark radiation, we get
\begin{equation}
    \frac{\rho_\text{GW}^\text{tot}}{\rho_\gamma}=\frac{7}{8}\left(\frac{4}{11}\right)^{4/3}\Delta N_\text{eff},
\end{equation}
where $\rho_\text{GW}^\text{tot}$ is the total energy density of GWs. The value of $\Delta N_\text{eff}$ is measured at the BBN time or the CMB time \footnote{The physical origins of these two methods are different and they generally give slightly different bounds \cite{Arbey:2021ysg}. The bound from CMB is generally stronger than the bound from BBN.}. But because GW and photons are redshifted in exactly the same way, the fraction of total energy density in GWs today is
\begin{equation}
    \Omega_\text{GW,0}^\text{tot}=\frac{\rho_{\gamma,0}}{\rho_\text{crit,0}}\frac{7}{8}\left(\frac{4}{11}\right)^{4/3}\Delta N_\text{eff}=1.15\times 10^{-5}\Delta N_\text{eff},
\end{equation}
where we used the current photon energy density $\rho_{\gamma,0}=2\pi^2T_0^4/30$ and the current critical energy density $\rho_\text{crit,0}=3H_0^2m_\text{pl}^2/8\pi$  with the Hubble constant $H_0\approx 67\ \rm km/s/Mpc=1.43\times 10^{-42}\ \rm GeV$ \cite{Planck:2018vyg}. By using $\Delta N_\text{eff}<0.3$ from Planck 2018 result at $95\%$ C.L. \cite{Planck:2018vyg}, we get
\begin{equation}\label{eq:Max of OmegaGW0total}
    \Omega_\text{GW,0}^\text{tot}<3.45\times 10^{-6}.
\end{equation}
This is a conservative bound. If there is also extra dark radiation in addition to GW, or if there are other primordial sources of GW, this upper bound could be reduced further as $\Delta N_\text{eff}$ encodes all forms of dark radiation \footnote{It should also be noted that GW, or dark radiation in general, produced after the CMB time is not subject to this bound. As an example, $\Omega_\text{GW}$ from supermassive BH mergers can exceed the $10^{-6}$ threshold as long as it does not violate any other late-time astrophysical constraints.}. If the GW spectrum is narrowly peaked, it is fine to just use the value of $\Omega_\text{GW}$ at the peak to compare with the above bound. In this paper, we adopt a more general formulation that adds up contributions of $\Omega_\text{GW}$ across frequencies:
\begin{equation}
    \Omega_\text{GW,0}^\text{tot}=\int_0^{k_\text{UV}}\Omega_\text{GW,0}(k)\frac{dk}{k},
\end{equation}
where $\Omega_\text{GW,0}(k)$ is given in Eq. \ref{eq:OmegaGW0}. Performing the integration, the upper bound from Eq. \ref{eq:Max of OmegaGW0total} gives an upper bound on RPBH population
\begin{equation}
    \beta<4.42\times 10^{-4}\gamma^{-1/2} A(l)^{17/12}B(l)^{17/24}\left(\frac{106.75}{g_{*,\text{eva}}}\right)^{11/96}\left(\frac{g_*(T_\text{H})}{106.75}\right)^{17/48}\left(\frac{g_{*,i}}{106.75}\right)^{17/96}\left(\frac{\rm g}{M_i}\right)^{17/24}.
\end{equation}
Together with the condition to have RPBH domination in Eq. \ref{eq:beta_c}, we obtain the following range:
\begin{equation}\label{eq:range of beta}
    1.78\times 10^{-6}\gamma^{-1/2}A(l)^2B(l)\left(\frac{\rm g}{M_i}\right)\lesssim \beta\lesssim 4.42\times 10^{-4}\gamma^{-1/2}A(l)^{17/12}B(l)^{17/24}\left(\frac{\rm g}{M_i}\right)^{17/24},
\end{equation}
where the $g_*$ factors have been chosen to take the standard normalized value $\sim 106.75$. If RPBH population is below this range, there is no RPBH domination and hence no isocurvature GW. If RPBH population is above this range, the isocurvature GW signal is too strong and is not allowed by $\Delta N_\text{eff}$ constraint.

\section{Results}\label{sec:Results}

After providing the generic framework to study isocurvature GW of ultra-light RPBHs, we now wish to work out an explicit example for illustration. We consider a modified version of the Simpson-Visser (SV) metric \cite{Simpson:2018tsi} where the lapse function is given by \cite{Loc:2025mzc}
\begin{equation}\label{eq:SV metric}
    f_\text{SV}(r)=g_\text{SV}(r)=1-\frac{2}{\sqrt{(r/GM)^2+l^2}},
\end{equation}
where $0<l<2$ is the dimensionless regularizing parameter. The curvature invariants and geodesics of this metric are regular at $r=0$. RBH with this metric does not form remnant but evaporates slowly toward zero mass. The evaporation process is self-similar, meaning that the semi-classical relations between BH's parameters are the same at every stage of evaporation similar to the behavior of singular Schwarzschild BH. The Hawking temperature and horizon radius, however, are different and are determined by the choice of $l$. The viewpoint that RBHs should evaporate completely was also supported in a separate study \cite{Khodadi:2025icd}.

The metric functions in Eq. \ref{eq:SV metric} give a smaller Hawking temperature $(A(l)<1)$ and a smaller horizon radius $(B(l)<1)$ than the Schwarzschild BH \cite{Loc:2025mzc}. According to Eq. \ref{eq:lifetime}, this implies an extended lifetime of RPBH. Intuitively, a lower Hawking temperature means that it takes longer for the black hole to lose all of its mass, and a smaller horizon size means that there are less ``doors" for the quanta to exit the event horizon so this factor also contributes to the extended lifetime. In the limit $l\rightarrow 0$, we have $\{A(l),B(l)\}\rightarrow 1$ and the usual singular Schwarzschild solution is recovered.

\begin{figure}[h!]
    \centering
    \includegraphics[width=\linewidth]{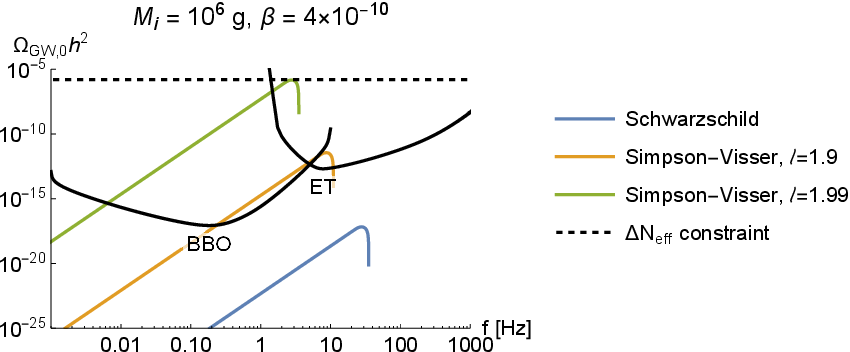}
    \caption{Isocurvature GW spectrum of Schwarzschild PBH and Simpson-Visser RPBH. The dashed black line indicates the upper bound on $\Omega_\text{GW,0}^\text{tot}$ coming from $\Delta N_\text{eff}$ constraint. The solid black contours are power-law integrated sensitivity curves of BBO and ET taken from \cite{Schmitz:2020syl}.}
    \label{fig:GW spectrum}
\end{figure}

By using Eq. \ref{eq:OmegaGW0}, we show in Fig. \ref{fig:GW spectrum} the isocurvature GW spectrum of PBHs for a given benchmark parameter $M_i=10^6\ \rm g$ and $\beta=4\times 10^{-10}$. The blue curve represents the singular Schwarzschild PBH case $(l=0)$. The orange $(l=1.9)$ and green $(l=1.99)$ curves represent the Simpson-Visser RPBHs. The close-to-extremal values of $l$ were chosen to highlight the difference between the two kinds of BH. The dashed black line indicates the absolute upper bound of the total GW spectrum given in Eq. \ref{eq:Max of OmegaGW0total}. The solid black contours are power-law integrated sensitivity curves of the corresponding GW detectors taken from \cite{Schmitz:2020syl}. We take $h=0.67$ to be the normalized Hubble constant.

From Fig. \ref{fig:GW spectrum}, we see that the peak frequency is shifted toward lower values for Simpson-Visser RPBHs.  This is because these RPBHs have extended lifetime and create a longer period of RPBH domination, leading to more redshift of frequency (see Eq. \ref{eq:kUV vs Mi}). For that same reason, the strength of isocurvature GW background is also enhanced as RPBHs' explosive decay would dramatically reheat the more subdominant radiation background (see Eq. \ref{eq:OmegaGW0}). More generally, the fact that a prolonged matter-dominated phase could enhance scalar-induced GW signal is also known in the literature \cite{Inomata:2019ivs}.

\begin{figure}[h!]
    \centering
    \includegraphics[width=\linewidth]{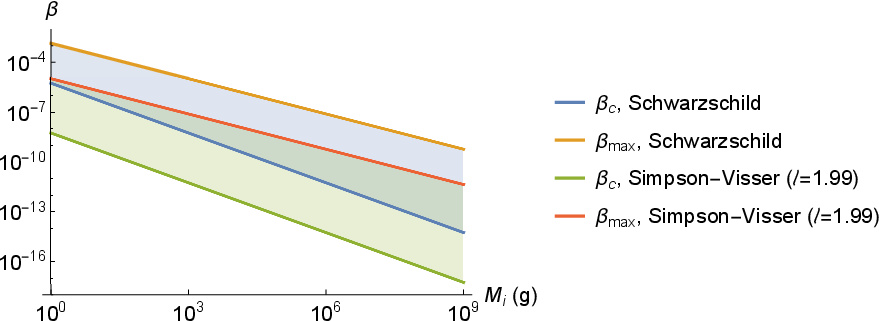}
    \caption{Allowed parameter space for the Schwarzschild PBH (shaded blue) and the Simpson-Visser RPBH (shaded green) where we can have isocurvature GW that is still consistent with $\Delta N_\text{eff}$ constraint. The parameter space below the $\beta_c$ line is \textit{not} ruled out; it is just where we don't have PBH domination and thus isocurvature GW. However, the parameter space above the $\beta_\text{max}$ line is \textit{absolutely} ruled out as there will be PBH domination with too strong isocurvature GW that is not permitted by $\Delta N_\text{eff}$ constraint.}
    \label{fig:population vs mass}
\end{figure}

We use Eq. \ref{eq:range of beta} to show in Fig. \ref{fig:population vs mass} the constraint on the population of PBHs as a function of their mass. The shaded blue region is allowed for Schwarzschild PBHs, whereas the shaded green region is allowed for the Simpson-Visser RPBHs with $l=1.99$. We should clarify that the parameter space below the line $\beta_c$ is \textit{not} ruled out. It is just the region where we don't have PBH domination and therefore there is no isocurvature GW that we discuss in this paper. However, the parameter space above the line $\beta_\text{max}$ is \textit{absolutely} ruled out as there will be definitely too strong isocurvature GW that is not permitted by $\Delta N_\text{eff}$ constraint.

From Fig. \ref{fig:population vs mass}, we see that the shaded region is shifted toward smaller $\beta$ for Simpson-Visser RPBHs. For a given initial mass, Simpson-Visser RPBHs have longer lifetime than Schwarzschild singular PBHs  (see Eq. \ref{eq:lifetime}) so that their population must be smaller in order to not have RPBH domination, leading to the reduced value of $\beta_c$. Once RPBHs dominate the Universe, their prolonged lifetime implies that their evaporation would reheat the more subdominant thermal bath more dramatically, leading to an enhanced isocurvature GW signal and thus the allowed maximum population $\beta{_\text{max}}$ must also decrease. The $\beta_c$ line is shifted more than the $\beta_\text{max}$ line as the former has stronger dependence on RPBH's temperature (characterized by $A(l)$) and horizon size (characterized by $B(l)$) than the latter (see Eq. \ref{eq:range of beta}).

\begin{figure}[h!]
    \centering
    \includegraphics[width=0.7\linewidth]{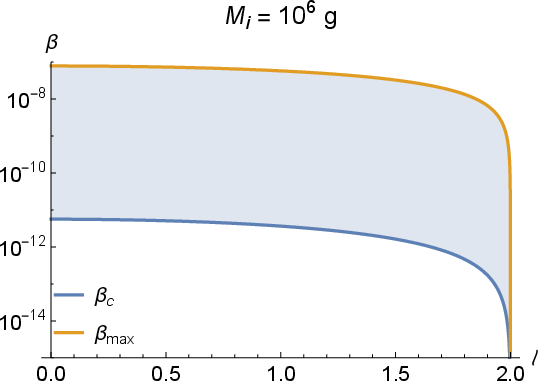}
    \caption{Constraints on the population PBH as a function of regularizing parameter $l$.}
    \label{fig:population vs l}
\end{figure}

Also by using Eq. \ref{eq:range of beta}, we can instead show in Fig. \ref{fig:population vs l} the constraints as a function of regularizing parameter $l$ for a fixed initial benchmark PBH mass $M_i=10^6\ \rm g$. From the figure, we see that the constraint decreases for increasing $l$ as the RPBHs live longer (see Fig. 2 of \cite{Loc:2025mzc}) and therefore produce stronger isocurvature GW. The singular Schwarzschild PBH case corresponds to $l=0$, whereas at $l=2$ the BH does not evaporate as there is no event horizon and the Hawking temperature is zero.

\section{Conclusion}\label{sec:Conclusion}

In this paper, we presented a scenario in which ultra-light RPBHs temporarily dominated the early Universe and subsequently evaporate to induce isocurvature GW. We derived constraints on the population of ultra-light RPBHs such that the total produced isocurvature GW does not violate $\Delta N_\text{eff}$ constraint on extra relativistic degrees of freedom. As example, we worked out explicitly the case of Simpson-Visser metric. RPBHs associated with this metric have lower Hawking temperature and smaller horizon size than the singular Schwarzschild PBHs, leading to extended lifetime of BH. The resulting isocurvature GW signal could be enhanced by orders of magnitude and therefore the resulting constraint on RPBH population could be much stronger. Comparing this result with the scenario of DM production from evaporation of RPBHs \cite{Loc:2025mzc}, a portion of the parameter space is now ruled out.

Our work could be extended in several ways. First, one can attempt to apply our generic formulae to other phenomenologically or theoretically motivated metrics to see how the results would change. Second, one may also study the effects of extended mass functions on the results. Third, one can consider the case of spinning RPBHs which is expected to alter BH's lifetime and hence the GW signal. We hope to explore these research avenues in future works.

\section*{Acknowledgments}
This work was supported by the National Science and Technology Council, the Ministry of Education (Higher Education Sprout Project NTU-114L104022-1), and the National Center for Theoretical Sciences of Taiwan.

\bibliographystyle{JHEP}
\bibliography{references}

\end{document}